\documentclass[aps,pre,twocolumn,showpacs,showkeys]{revtex4}
\usepackage{epsfig}
\usepackage{times}
\usepackage{amsmath}
\begin{document}

\title{Synchronization transitions on scale-free neuronal networks due to finite information transmission delays}

\author{Qingyun Wang,$^{\star,\dagger}$ Matja{\v z} Perc,$^\ddagger$ Zhisheng Duan,$^\star$ and Guanrong Chen$^{\star,\S}$}

\affiliation
{$^\star$State Key Laboratory for Turbulence and Complex Systems, Department of Mechanics and Aerospace Engineering, College of Engineering, Peking University, Beijing 100871, China \\
$^\dagger$School of Statistics and Mathematics, Inner Mongolia Finance and Economics College, Huhhot 010051, China \\
$^\ddagger$Department of Physics, Faculty of Natural Sciences and Mathematics, University of \\ Maribor, Koro{\v s}ka cesta 160, SI-2000 Maribor, Slovenia \\
$^\S$Department of Electronic Engineering, City University of Hong Kong, Hong Kong SAR, China}

\begin{abstract}
We investigate front propagation and synchronization transitions in dependence on the information transmission delay and coupling strength over scale-free neuronal networks with different average degrees and scaling exponents. As the underlying model of neuronal dynamics, we use the efficient Rulkov map with additive noise. We show that increasing the coupling strength enhances synchronization monotonously, whereas delay plays a more subtle role. In particular, we found that depending on the inherent oscillation frequency of individual neurons, regions of irregular and regular propagating excitatory fronts appear intermittently as the delay increases. These delay-induced synchronization transitions manifest as well-expressed minima in the measure for spatial synchrony, appearing at every multiple of the oscillation frequency. Larger coupling strengths or average degrees can broaden the region of regular propagating fronts by a given information transmission delay and further improve synchronization. These results are robust against variations in system size, intensity of additive noise and the scaling exponent of the underlying scale-free topology. We argue that fine-tuned information transmission delays are vital for assuring optimally synchronized excitatory fronts on complex neuronal networks, and indeed, they should be seen as important as the coupling strength or the overall density of interneuronal connections. We finally discuss some biological implications of the presented results.
\end{abstract}

\maketitle

\section{Introduction}

Synchronization phenomena are ubiquitous in nature and play an important role in biology, ecology, climatology, sociology, technology, and even fine arts \cite{a1, a2, a3, a4}. In the study of nonlinear dynamical systems, synchronization is recurrently being placed in the focus of attention, and recently insightful findings regarding the synchronization phenomena on complex networks were reported \cite{add1, add2, add3, add4, add5, addc} and comprehensively reviewed \cite{a3}. It is interesting to see that one can literary infer topological scales of complex networks based solely on synchronization \cite{add6}, thus making a closed loop of dependence between the synchronizability and the structure of underlying interactions of network elements. In neural systems, in particular, the interplay between the network structure and the dynamics taking place on it is closely interrelated. The function-follow-form paradigm, for example, is central to attest to this observation \cite{ff1, ff2, ff3, ff4, ff5}. It is also well known that the cerebral cortex features properties that are characteristic for complex networks \cite{cn1, cn2, cn3}. As a results, the firing activities of individual neurons are often related to the synchronization of the underlying network, and accordingly synchronized firings can be observed at virtually all processing levels, including the retina \cite{ns1, ns2}, the lateral geniculate nucleus \cite{ns2}, and the cortex \cite{ns3, ns4, ns5, ns6}.

Synchronization on complex networks, therefore, has actually become a focal topic in theoretical neurosciences \cite{r1, r2}, as evidenced by several recent studies devoted to the explorations of this subject \cite{fh0, fh1, qy1, qy2, fh2, fh3, fh4, fh5, fh6}. Important works have been elaborating on general aspects of synchronization on scale-free \cite{ua1, ua2, uam, uay} as well as complex-gradient \cite{ua3} networks, among many other models. It is now clear that synchronization is key to the efficient processing and transmission of information across a nervous system such as the brain \cite{ns4, ns7}. The handling of information transmission over a neuronal network, however, is still an open avenue for research. Since information transmission delays are inherent to neuronal systems because of the finite speeds at which action potentials propagate across neuron axons, and due to the time lapses occurring in both dendritic and synaptic processing \cite{or23}, studies are in need of catching up with the most recent advances in synchronization research on complex networks.

Indeed, delays have been found responsible for several interesting phenomena in coupled dynamical systems. For example, Ernst \textit{et al.} \cite{ernst1} have identified mechanisms of synchronization among pulse-coupled oscillators in the presence of time delay. Moreover, it has been shown that coupled oscillators undergo a transition towards amplitude death faster if the time delays in coupling are distributed over an interval rather than being uniform throughout the system \cite{atadd1}. The role of delays and connection topologies for the synchronization of coupled chaotic maps has been studied in \cite{axf1}, where it is reported that on scale-free and random networks sufficiently large coupling strengths can offset the delayed flow of information. It has also been shown that networks with delays can sometimes synchronize more easily than in their absence, and it has been argued that this may be particularly relevant for neuronal networks for establishing a concept of collective information processing in the presence of delayed information transmission \cite{gerst1}. In the present paper we give further support to the latter assumption by considering the impact of delays in scale-free neuronal networks. More recently, the role of delays by the formation of the so-called chimera, \textit{i.e.} coexistent coherent and incoherent states in a system of nonlocally coupled phase oscillators has also been examined \cite{seth1}, and it has been shown that time-delays can induce a transition towards phase clustering, giving rise to clustered chimera states that have spatially distributed phase coherence separated by incoherence with adjacent coherent regions in antiphase. Here we support the theory of delay-induced dynamical transitions in terms of the synchrony of neuronal noise-induced excitations on a scale-free network. Notably, previous studies have already considered particularly the neuronal dynamics on large networks in conjunction with information transmission delay, but the focus was primarily on the bifurcation structure of transitions between different delay-induced states, including oscillatory bumps, aperiodic regimes, traveling, lurching and standing waves, as well as regimes of multistability \cite{AR1, AR2}. The impact of information transmission delay on neuronal synchronization, on the other hand, has been studied in \cite{or24, or25, or26}, where the emergence of zigzag fronts, clustering antiphase synchronization and in-phase synchronization on regular and small-world neuronal networks has been discussed.

At present, we aim to extend the scope of the above-mentioned investigations by studying front propagation and synchronization transitions in dependence on the coupling strength and information transmission delays over scale-free neuronal networks with different average degrees and scaling exponents. Notably, it has been reported that, by using functional magnetic resonance imaging, power-law distributions can be obtained upon linking correlated fMRI voxels \cite{cn2}, and that the robustness against simulated lesions of anatomic cortical networks relies mostly on the scale-free structure \cite{sm2}. This study thus addresses a relevant system setup that is still widely open for new research. More specifically, we report several non-trivial effects induced by finite delay lengths, and the ability of its fine-tuning towards highly synchronized fronts of excitations. These findings are compared to the impacts of different coupling strengths and average degrees, and their robustness is examined at different levels of additive noise, variations in system size and different scaling exponents of the underlying scale-free topology. Remarkably, we found that, irrespective of the system size and the scaling exponent, properly adjusted information transmission delays play a pivotal role in warranting synchronized fronts of excitations on noisy scale-free neuronal networks, which can be further enhanced via larger coupling strengths or higher average degrees of the constitutive nodes. We argue that this is primarily attributed to the emergence of locking between the delay and the inherent oscillation frequency of individual neurons of the scale-free network.

The remainder of this paper is organized as follows. In the next section, we describe the Rulkov map \cite{r20}, which will be employed to obtain an efficient setup for simulating neuronal dynamics on scale-free networks \cite{r21}. In Section II we also present the coupling scheme and the measure for synchronization of excitatory fronts, as well as other mathematical methods to be used. In Section III we present the main results, and in the last Section we summarize our findings and discuss their potential implications.

\section{Mathematical model and setup}

For simulating the neuronal dynamics on a scale-free network effectively, the Rulkov map \cite{r20} is employed, which succinctly captures all the major
dynamical features of the complex continuous-time models. The spatial-temporal evolution of the studied network, corrupted with additive Gaussian noise and experienced with information transmission delays, is described by the following iteration equations
\begin{eqnarray}
x^{(i)}(n+1)&=& \alpha f[x^{(i)}(n)]+y^{(i)}(n)+w
\xi^{(i)}(n)\nonumber\\*
&&+ D\sum_{j}\varepsilon^{{i,j}}\left[x^{j}(n-\tau)-x^{i}(n)\right], \\
y^{(i)}(n+1)&=& y^{(i)}(n)-\beta x^{(i)}(n)-\gamma, \ i=1,\ldots,N
\nonumber
\end{eqnarray}
where $n$ is the discrete time index, $x^{(i)}(n)$ is the membrane potential and $y^{(i)}(n)$ the variation of ion concentration of the $i$-th neuron, representing the fast and the slow variable, respectively. The slow temporal evolution of $y^{(i)}(n)$ is due to the small values of the positive parameters $\beta$ and $\gamma$, which within this study equal $\beta = \gamma = 0.001$ unless stated otherwise. Moreover, $\alpha$ is the main parameter
determining the dynamics of individual neurons on the scale-free network. According to \cite{r20}, if $\alpha < 2.0$ all neurons are situated in excitable steady states $[x^* = -1,y^* = -1-(\alpha/2)]$, whereas if $\alpha > 2.0$ complex firing and bursting patterns of temporal activity emerge via a Hopf bifurcation. Here we set $\alpha=1.95$ and initiate each neuron from steady state initial conditions, so that the additive spatiotemporal Gaussian noise $\xi_{i}(n)$, having mean $<\xi_{i}>=0$ and autocorrelation $<\xi_{i}(n) \xi_{j}(h)>=\delta_{ij}\delta(n-h)$, acts as the only source of large-amplitude excitations. Moreover, in Eq.~(1) $f(x)=\frac{1}{1+x^2}$ is a nonlinear function warranting the essential ingredient of neuronal dynamics, parameter $w$ determines the noise intensity, $D$ is the coupling strength, and $\tau$ is the information transmission delay. The latter two parameters will be in the focus of attention within this work, whereas $w$, $\beta$ and $\gamma$ will be varied only occasionally.

\begin{figure*}
\centerline{\epsfig{file=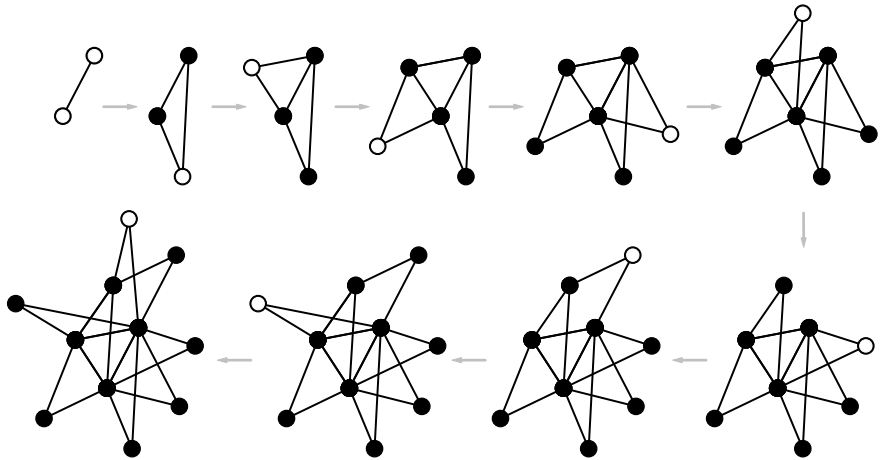,width=13cm}}
\caption{\label{fig1} Schematic presentation of growth and preferential attachment, as proposed by Barab\'{a}si and Albert \cite{r21}. Each new node (white) preferentially attaches to two (thus here $k_{av}=4$) old nodes (black) that already have many other connections at that time.}
\end{figure*}

As the interaction base between neurons we primarily use the scale-free network generated via growth and preferential attachment as proposed by Barab\'{a}si and Albert \cite{r21}, consisting of $N=200$ nodes unless stated otherwise. Each node corresponds to one neuron, whose dynamics is governed by the noise-driven Rulkov map. In Eq.~(1) $\varepsilon^{i,j} = 1$ if neuron $i$ is coupled to neuron $j$ and $\varepsilon^{i,j} = 0$ otherwise. Following ref. \cite{r21}, the preferential attachment is introduced via the probability $\Pi$, which states that a new node will be connected to node $i$ depending on its connectivity $k_i$ according to $\Pi(k_i )=k_i /\sum_jk_j$, as demonstrated schematically in Fig.~\ref{fig1}. This growth and preferential attachment scheme yields a network with an average degree $k_{av}=\frac{\sum_i k_i}{N}$, and a power-law degree distribution with the slope of the line equaling $\approx -3$ on a double-logarithmic graph. Notably, analytical estimations predict the slope of the line to equal $-3$ \cite{r21}. We will use Barab\'{a}si-Albert scale-free networks having $k_{av}=4$ throughout this work (see Fig.~\ref{fig1}), unless stated otherwise.

In order to quantitatively study the degree of spatiotemporal synchronization in the network, and thus support below presented visual assessments of front propagation via space-time plots, we introduce, by means of the standard deviation, a synchronization parameter $\sigma$ (see \textit{e.g.} \cite{r22}), which can be calculated effectively according to:
\begin{equation}
\sigma=\frac{1}{T}\sum\limits_{n=1}^{T}\sigma(n), \ \
\sigma(n)=\frac{1}{N}\sum\limits_{i=1}^{N}[x^{i}(n)]^2-[\frac{1}{N}\sum\limits_{i=1}^{N}x^{i}(n)]^2
\end{equation}
It turns out that $\sigma$ is an excellent indicator for numerically measuring the spatiotemporal synchronization of excitations, hence revealing different synchronization levels and related transitions. From Eq.~(2), it is evident that the smaller the synchronization parameter $\sigma$, the more synchronous the neuronal network. Accordingly, when $\sigma=0$ the network reaches complete synchrony. Final results shown below were averaged over $20$ independent runs for each set of parameter values to warrant appropriate statistical accuracy with respect to the scale-free network generation and numerical simulations.

\section{Results}

\begin{figure*}
\centerline{\epsfig{file=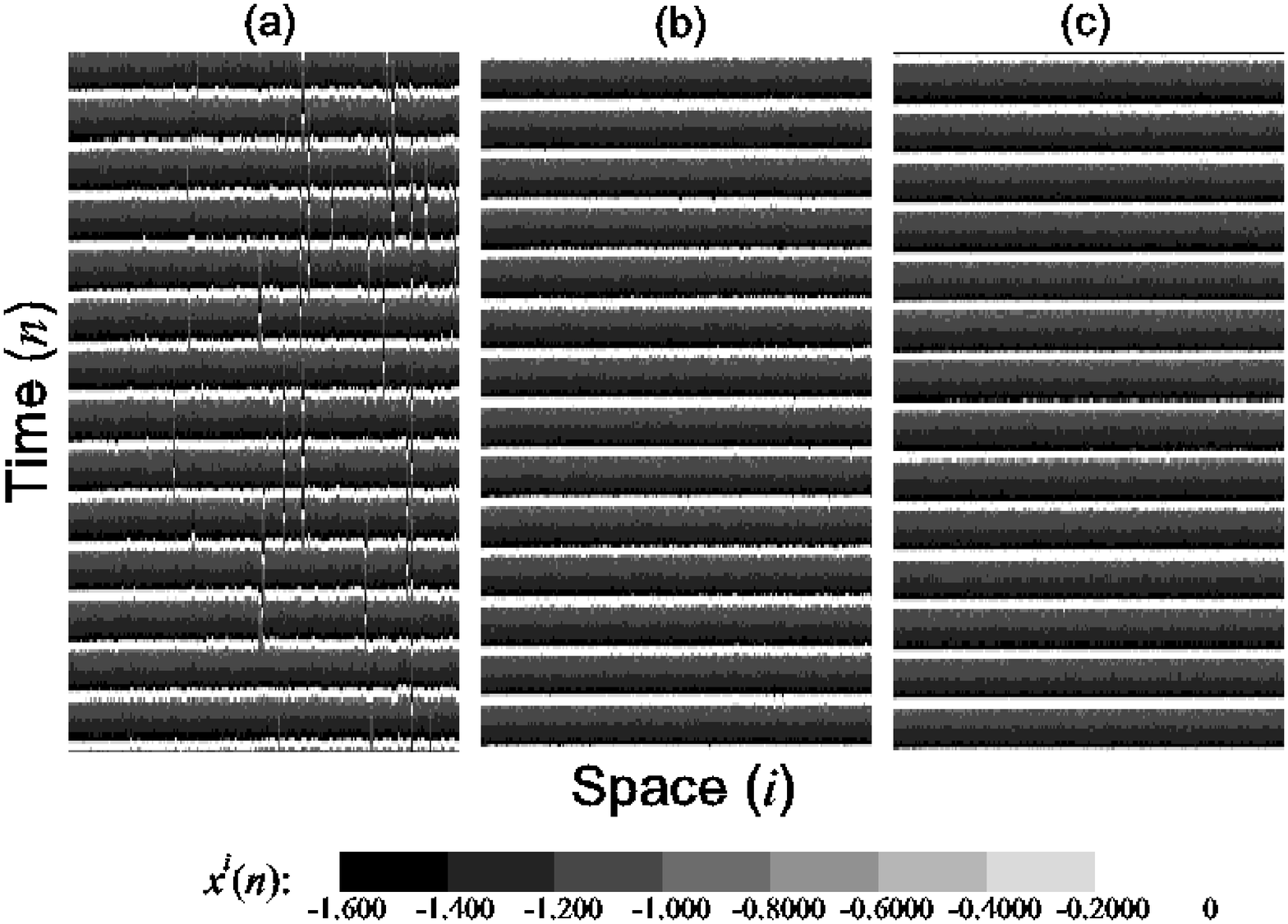,width=7cm} \ \epsfig{file=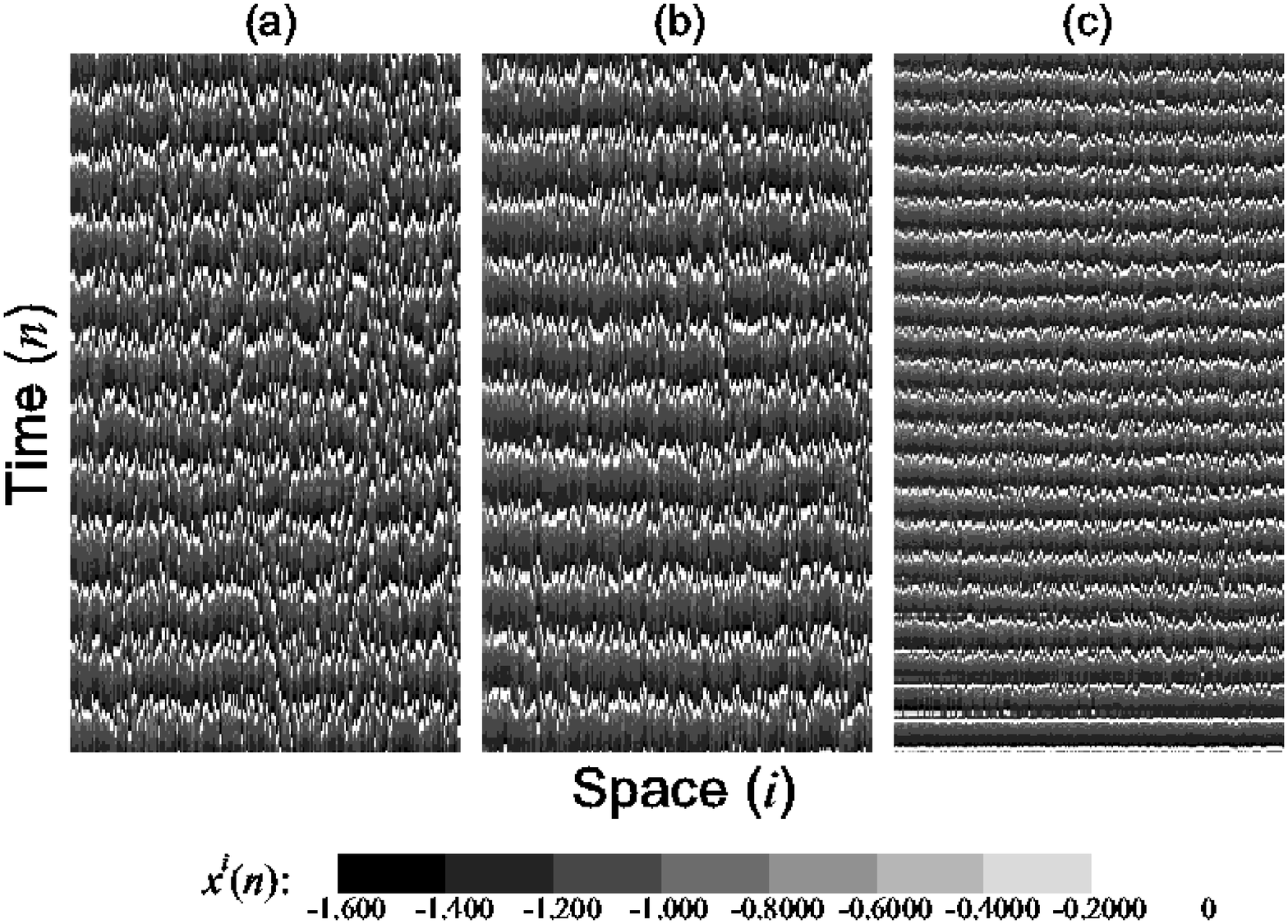,width=7cm}}
\caption{\label{fig2} Left: Space-time plots of $x^{(i)}(n)$ obtained for $\tau=700$ and $w=0.015$ with different coupling strengths $D$, equaling: (a) $0.004$, (b) $0.008$, (c) $0.016$. Right: Space-time plots of $x^{(i)}(n)$ obtained for $\tau=1000$ and $w=0.015$ with different coupling strengths $D$, equaling: (a) $0.004$, (b) $0.008$, (c) $0.016$. In all panels, the system size is $i=1,2 \dots, 200=N$.}
\end{figure*}

We start by presenting space-time plots obtained with a fixed information transmission delay $\tau=700$ and noise intensity $w=0.015$, but different values of the coupling strength $D$. Results shown in the left three panels of Fig.~\ref{fig2} illustrate the spatiotemporal dynamics of neurons on the scale-free neuronal network having $k_{av}=4$. Evidently, for small coupling strengths [see panel (a), left] the excitatory fronts are quite nicely ordered in both time and space. However, both the temporal and spatial regularity increase further and substantially as $D$ is increased [see panels (b) and (c), left]. Interestingly, by setting the information transmission delay to $\tau=1000$ and keeping the same noise intensity $w=0.015$, the order in both time and space deteriorates substantially, as depicted by the right three panels of Fig.~\ref{fig2}. Nevertheless, increasing the coupling strength can still improve the overall regularity of the excitatory fronts [comparing panels (a), (b) and (c), right]. It is thus revealed that different information transmission delays have a profound impact on the spatiotemporal regularity of excitatory fronts, whereas increasing the coupling strength always leads to an improvement of temporal and spatial synchronization.

\begin{figure}
\centerline{\epsfig{file=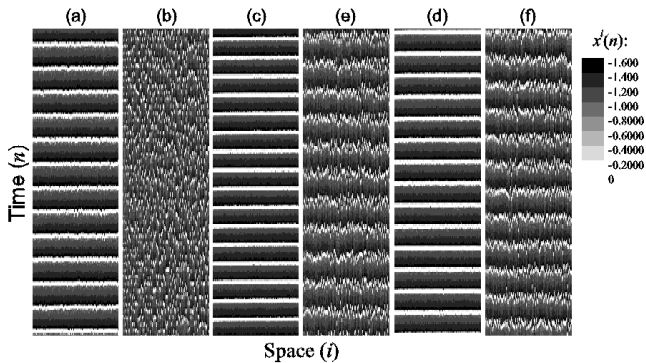,width=8.5cm}}
\caption{\label{fig3} Space-time plots of $x^{(i)}(n)$ obtained for $D=0.01$ and $w=0.015$ with different information transmission delays $\tau$, equalling: (a) $0$, (b) $200$, (c) $600$, (d) $1000$, (e) $1400$, (f) $1800$. In all panels the system size is $i=1,2 \dots, 200=N$.}
\end{figure}

To investigate the impact of different information transmission delays, outlined in Fig.~\ref{fig2} more precisely, we show in Fig.~\ref{fig3} space-time plots obtained with fixed coupling strength $D=0.01$ and noise intensity $w=0.015$, but different values of $\tau$. It can be observed that the spatiotemporal dynamics is ordered nicely if $\tau=0$ [see panel (a)]. When $\tau=200$, this deteriorates drastically [see panel (b)], but is again revived at $\tau=600$ [see panel (c)]. In fact, by closely examining space-time plots obtained with $\tau=0$ and $\tau=600$, respectively, one can observe that the non-zero yet appropriately tuned information transmission delay can further enhance the regularity of excitatory fronts as compared to the case of $\tau=0$. Quite remarkably, when $\tau=1000$ the regularity of excitatory fronts is again heavily impaired [see panel (d)], yet with $\tau=1400$ the order is restored anew [see panel (e)]. Indeed, the information transmission delay induced transitions to superbly synchronized neuronal activities on scale-free networks seem to appear intermittently, at roughly integer multiples of a given value of $\tau$, which equals approximately $600-700$ in Fig.~\ref{fig3}. In accordance with this preliminary assessment, it is expectable that with $\tau=1800$ disorder in the temporal as well as the spatial domain sets in again [see panel (f)]. Visual investigations of Fig.~\ref{fig3} thus reveal that regular and irregular front propagation appears intermittently as the delay is increased. Hence, it can be stated that finite (non-zero) information transmission delays play a pivotal role in the generation of spatiotemporal patterns of neuronal activity on scale-free networks.

\begin{figure*}
\centerline{\epsfig{file=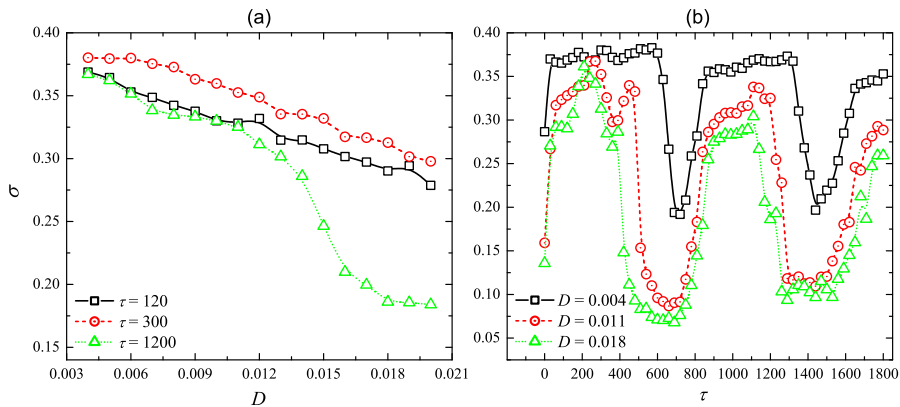,width=12cm}}
\caption{\label{fig4} (color online) (a) Dependence of the synchronization parameter $\sigma$ on $D$ with different $\tau$. (b) Dependence of $\sigma$ on $\tau$ with different $D$. Where applicable, other parameters are the same as in Fig.~\ref{fig2}.}
\end{figure*}

In what follows, the degree of spatiotemporal synchronization will be studied quantitatively via $\sigma$ (see Eq.~2), so as to support and validate the above visual assessments. Furthermore, it remains of interest to examine the impact of different $w$ and $k_{av}$. In Fig.~\ref{fig4}(a), we first plot $\sigma$ in dependence on $D$ for three different values of $\tau$. As visually interpreted by space-time plots presented in Fig.~\ref{fig2}, larger coupling strengths indeed facilitate spatiotemporal synchronization in an monotonous manner. That is, as $D$ increases, $\sigma$ decreases (irrespective of $\tau$), which is in agreement with previous studies examining synchronization phenomena in neuronal as well as many other nonlinear systems. More eventful are results presented in Fig.~\ref{fig4}(b), where $\sigma$ is presented in dependence on $\tau$ for three different coupling strengths $D$. It can be observed clearly that certain values of $\tau$ significantly facilitate spatiotemporal synchronization of excitatory fronts on scale-free neuronal networks. The two minima of $\sigma$ appear at $\tau \approx 700$ and $\tau \approx 1400$, respectively, and are largely independent of $D$. This confirms the above claim that the information transmission delay induced transitions to spatiotemporally synchronized neuronal activity appear intermittently, at integer multiples of the given value of $\tau$. On the other hand, values of $\tau$ outside these regions significantly impair synchronization, as can be inferred from the rather sharp ascends towards larger $\sigma$ beyond the optimal delays.

\begin{figure*}
\centerline{\epsfig{file=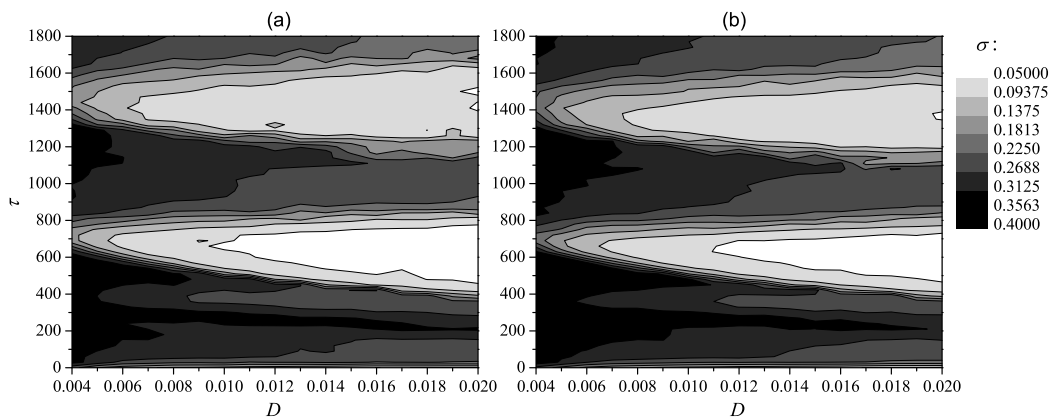,width=14cm}}
\caption{\label{fig5} Contour plots of $\sigma$ in dependence on $D$ and $\tau$ with: (a) $w=0.015$ and (b) $w=0.03$. Information transmission delay induced synchronization transitions are clearly visible and largely robust to variations of the noise intensity $w$.}
\end{figure*}

The delay-induced synchronization transitions, as well as the impact of increasing $D$, can be presented succinctly also via contour plots of $\sigma$ in dependence on the two parameters. Figure~\ref{fig5} features two such graphs obtained for $w=0.015$ [panel (a)] and $w=0.03$ [panel (b)]. Transitions to highly synchronized states in dependence on the information transition delay are clearly visible as extensive white regions (denoting smallest values of $\sigma$) occurring at $\tau \approx 700$ and $\tau \approx 1400$, corresponding to the two minima depicted previously in Fig.~\ref{fig4}(b). Moreover, results presented in Fig.~\ref{fig5} clearly convey the impact of increasing the coupling strength $D$. In fact, not only do increasing $D$ decrease $\sigma$, as outlined above when interpreting Fig.~\ref{fig4}(a), but also they broaden the span of $\tau$ within which synchronous spatiotemporal neuronal activity is warranted. Notably, the optimal values of $\tau$ shift insignificantly during the broadening. These features can be inferred also from Fig.~\ref{fig4}(b), yet the contour plots in Fig.~\ref{fig5} convey them more clearly. Finally, it is interesting to note that different values of $w$ do not evoke significantly different results, as can be appreciated by comparing panels (a) and (b) of Fig.~\ref{fig5}. From this, we conclude that the delay-induced transitions to synchronous neuronal activity on scale-free networks are largely independent of $D$ (apart from the broadening of the interval of $\tau$ warranting optimal synchronization) and robust against reasonable variations of the noise intensity.

\begin{figure*}
\centerline{\epsfig{file=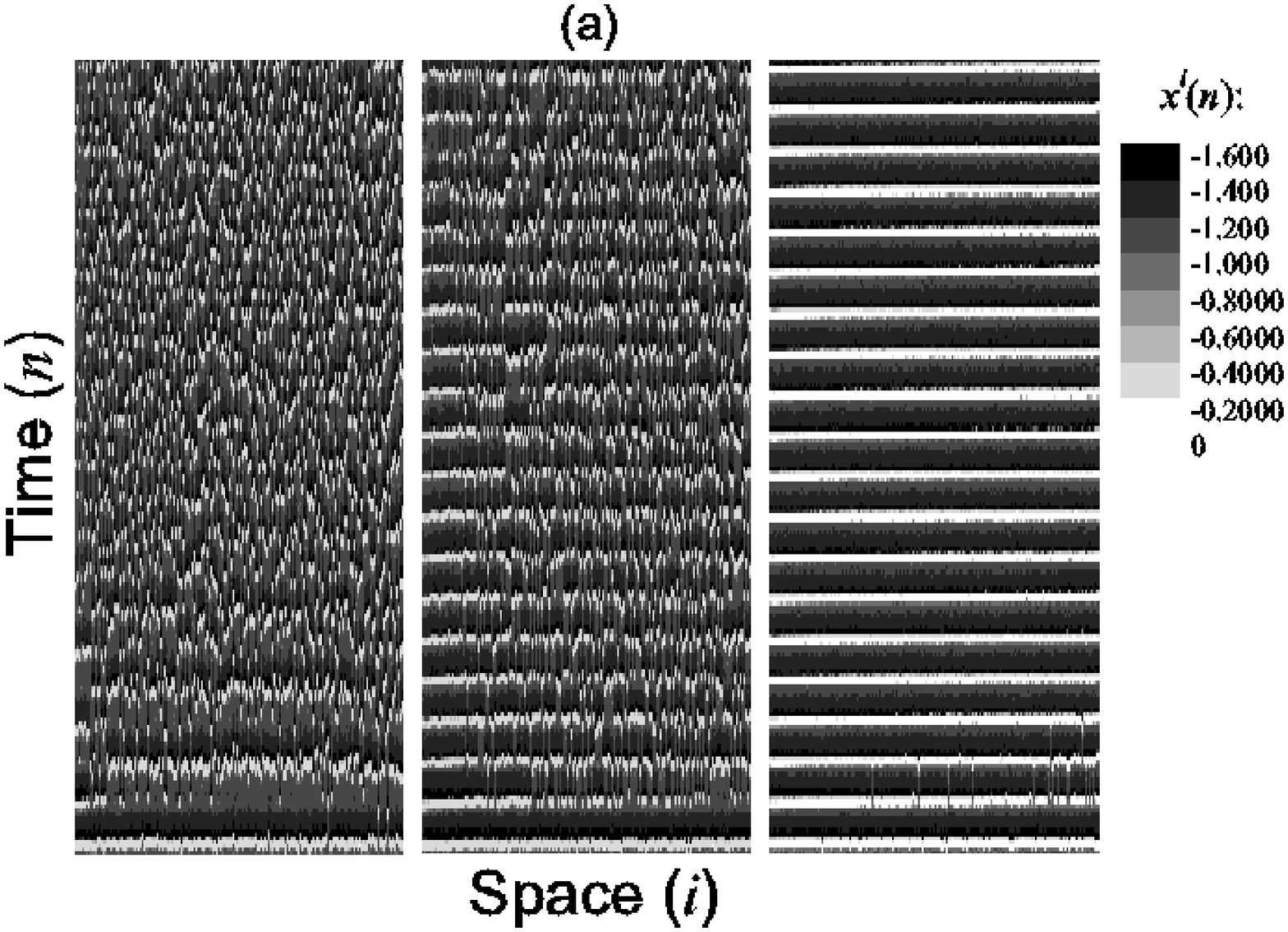,width=7cm}\epsfig{file=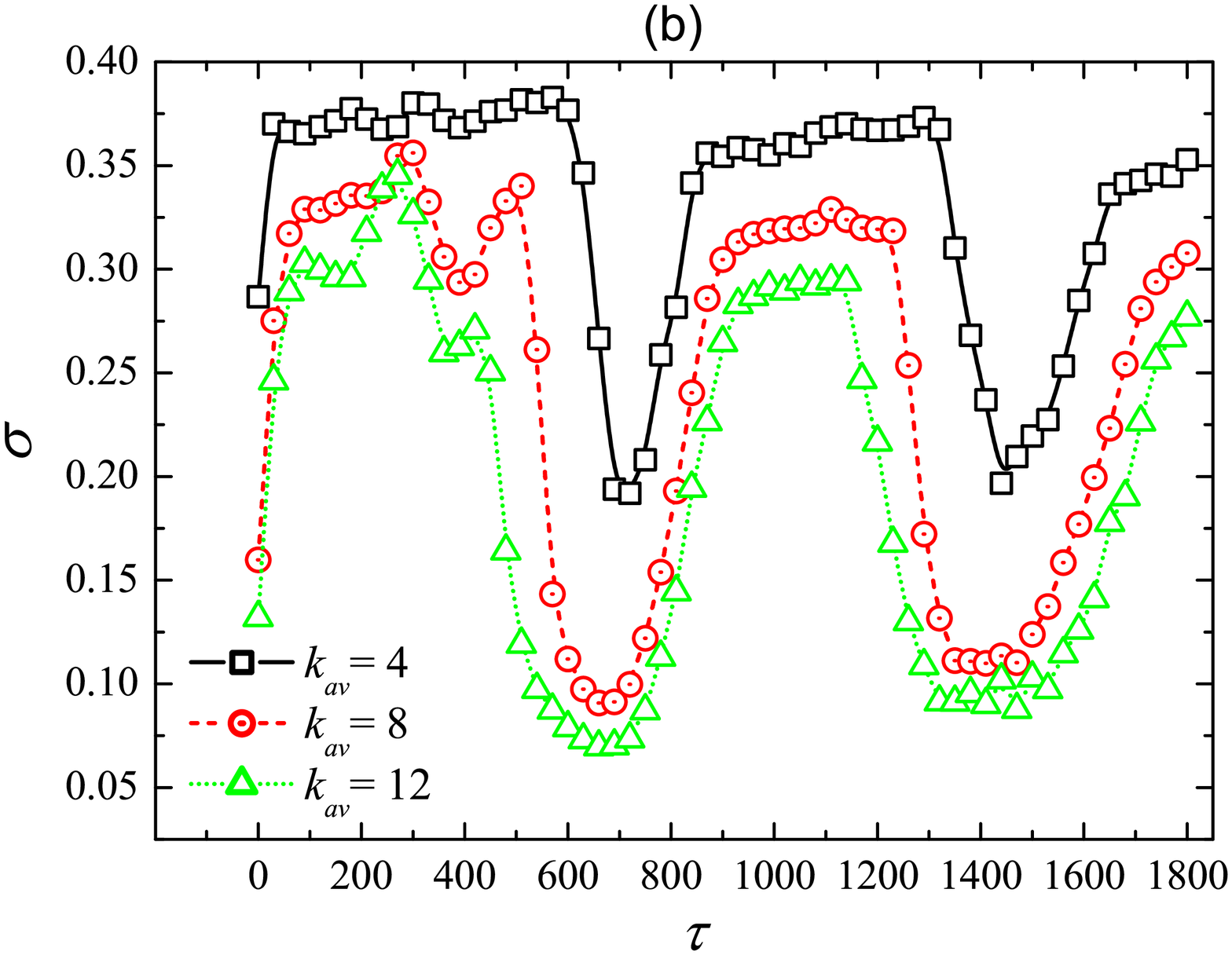,width=6.3cm}}
\caption{\label{fig6} (color online) (a) Space-time plots of $x^{(i)}(n)$ obtained with $D=0.004$, $w=0.015$ and $\tau=500$ for different average degrees $k_{av}$, equalling (from left to right) $4$, $8$ and $12$, respectively. In all presented panels the system size is $i=1,2 \dots, 200=N$. (b) Dependence of $\sigma$ on $\tau$ with different $k_{av}$. Other parameter values are the same as in panel (a).}
\end{figure*}

Thus far, we have considered only scale-free networks having $k_{av}=4$. Since the average degree determines the density of interneuronal links, and is thus arguably an important parameter, we present effects of different $k_{av}$ in Fig.~\ref{fig6}. Panel (a) features space-time plots obtained with $D=0.004$ and $\tau=500$ for increasing average degree from left to right. Visual inspection reveals that the impact of increasing $k_{av}$ is comparable to the impact of increasing $D$ (comparing with the space-time plots presented in Fig.~\ref{fig2}) in that the excitatory fronts propagate increasingly ordered in both time and space as the average degree increases. This can be confirmed quantitatively via $\sigma$, as shown in Fig.~\ref{fig6}(b). Indeed, larger $k_{av}$ shift lower the whole outlay of $\sigma$ in dependence on $\tau$, thus indicating improvement in the quality of spatiotemporal synchronization of neuronal activity on scale-free networks. Importantly, however, the oscillating outlay of $\sigma$, along with the optimal value of $\tau$, is preserved and does not vary in dependence on $k_{av}$. Therefore we conclude that the delay-induced synchronization transitions are a robust phenomenon of neuronal dynamics on scale-free networks, which an important role in achieving synchronized information transmission among neighboring as well as distant neurons.

\begin{figure*}
\centerline{\epsfig{file=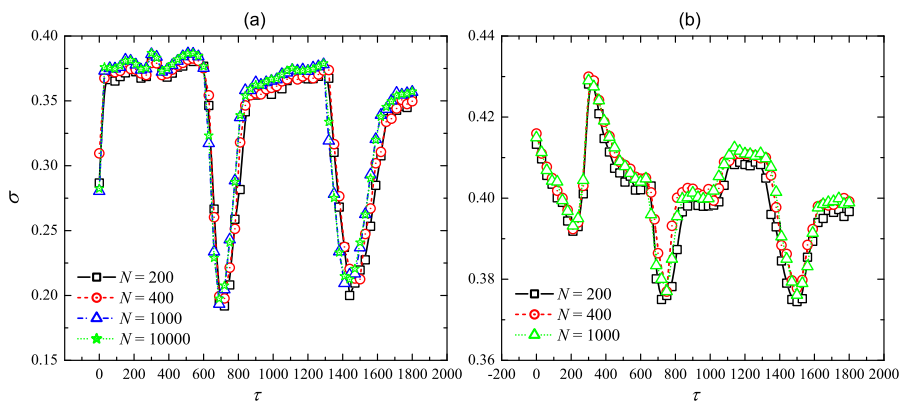,width=12cm}}
\caption{\label{fig7} (color online) (a) Dependence of $\sigma$ on $\tau$ with different $N$ of the Barab\'{a}si-Albert scale-free network \cite{r21} having $k_{av}=4$. Other parameter values are $D=0.004$ and $w=0.015$. (b) Dependence of $\sigma$ on $\tau$ with different $N$ of the scale-free network generated as proposed in \cite{sfax} (see also text for details). Other parameter values are $D=0.008$ and $w=0.015$.}
\end{figure*}

In order to further test the generality of our findings, we examine the impact of different $N$ as well as different scaling exponents characterizing the scale-free degree distribution. While the growth and preferential attachment algorithm proposed by Barab\'{a}si and Albert \cite{r21} yields a power-law degree distribution with the slope of the line $\approx -3$ on a double-logarithmic graph and serves as the most commonly used structure for testing theoretical models, also relevant for neuronal networks are slopes around $-2$, as reported in \cite{cn2}. Accordingly, we employ an alternative algorithm for scale-free network generation based on assigning a quenched fitness value to every node, and drawing links among them with a probability depending on the fitness of the two involved sites \cite{sfax}. Using exponentially distributed fitness and a threshold rule for linking nodes, we obtain a scale-free network with the scaling exponent $-2$, as presented in Fig. 3 of \cite{sfax}. Results presented in Figs.~\ref{fig7}(a) and (b) show clearly that variations of the system size do not notably influence the outcome of our simulations. In fact, the minima of $\sigma$ remain located at the same values of $\tau$ and are of the same depth (with a reasonable error margin) irrespective of $N$. Likewise, by changing the scaling exponent from $-3$ (as given by the Barab\'{a}si-Albert algorithm) to $-2$ (as given by the algorithm described in \cite{sfax}) the results also do not change significantly in that the minima of $\sigma$ appear by roughly the same values of $\tau$ [compare panels (a) and (b) of Fig.~\ref{fig7}]. Notable as a qualitative difference between the two scaling exponents characterizing the underlying scale-free topology is the very first local minimum of $\sigma$ by $\tau \approx 250$, occurring if the slope of the degree distribution equals $-2$, as depicted in Fig.~\ref{fig7}(b). We conjecture that this minimum may be related to the subharmonic of the optimal $\tau$ ($1/3$ of the first minimum at $\tau \approx 740$). Nevertheless, results presented in Fig.~\ref{fig7} attest to the fact that reported synchronization transitions on scale-free neuronal networks due to finite information transmission delays are largely robust to variations of the system size and of the scaling exponent characterizing the scale-free degree distribution.

\begin{figure}
\centerline{\epsfig{file=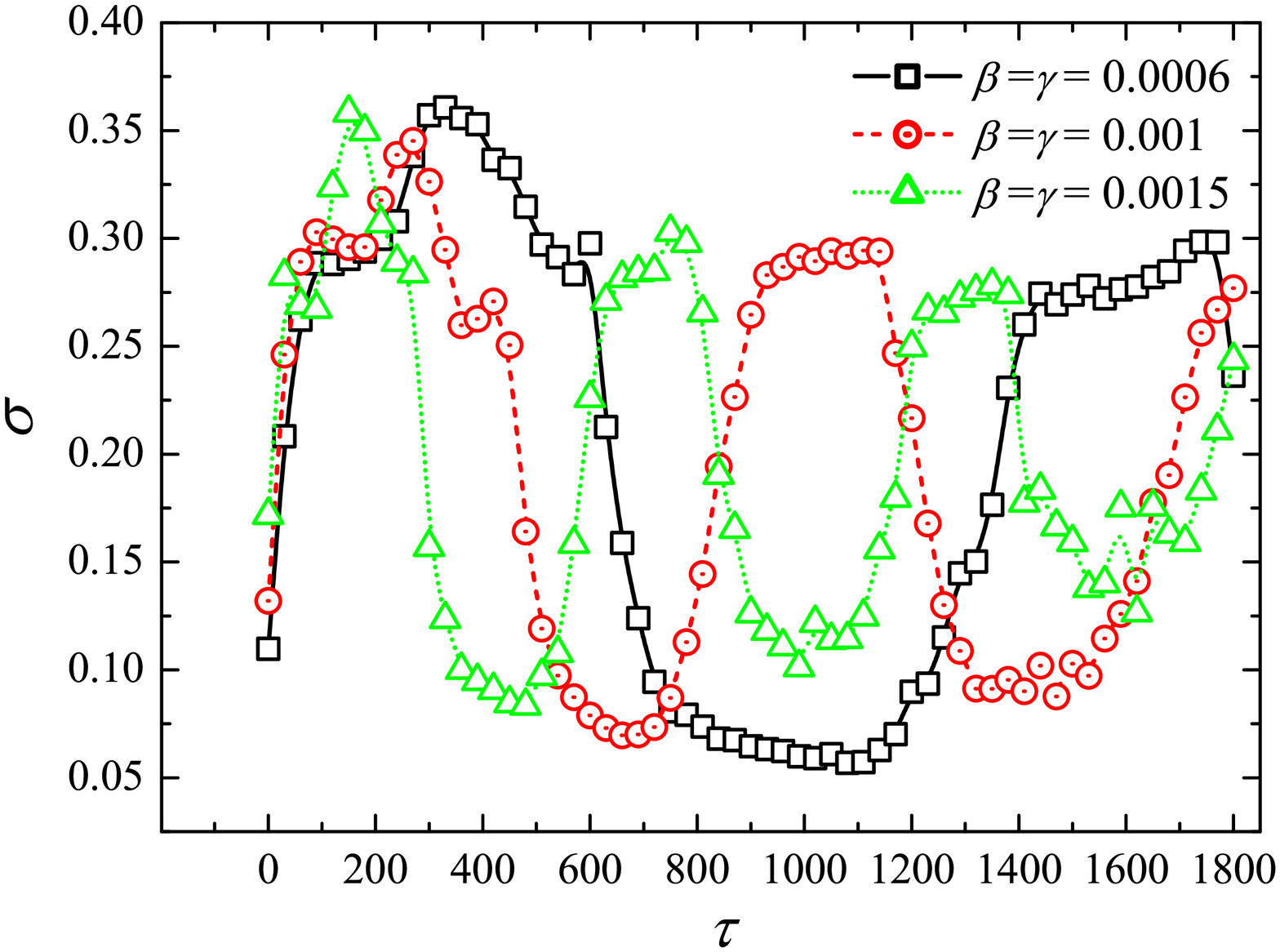,width=6.5cm}}
\caption{\label{fig8} (color online) Dependence of $\sigma$ on $\tau$ for different combinations of $\beta$ and $\gamma$. Other parameter values are: $D=0.018$, $w=0.015$ and $k_{av}=4$ (Barab\'{a}si-Albert scale-free network \cite{r21}).}
\end{figure}

Finally in this section, we provide an explanation for the emergence of the newly reported delay-induced synchronization transitions. Up to now we have shown that the optimal value of $\tau$, resulting in the occurrence of the first minimum of $\sigma$, as well as its reappearance at integer multiples, does not vary significantly in dependence on $D$ [see Fig.~\ref{fig4}(b)], $w$ (see Fig.~\ref{fig5}), $k_{av}$ [see Fig.~\ref{fig6}(b)], $N$ [see Fig.~\ref{fig7}(a)] or the scaling exponent characterizing the underlying scale-free network [see Fig.~\ref{fig7}(b)]. This leads to the conclusion that parameters determining the global neuronal dynamics do not play a significant role. Hence, one may adjust the local dynamics of each neuron by varying $\beta$ and $\gamma$ (thus far we have not varied them). These two parameters affect the speed of the temporal evolution of $y_{i}(n)$, and consequently the predominant oscillation period of excitations. Results presented in Fig.~\ref{fig8} clearly show that the locations of minima of $\sigma$ shift to different values of $\tau$ as $\beta$ and $\gamma$ are varied. We thus determined the predominant oscillation period $t_{osc}$ of individual neurons within the scale-free network, and remarkably found that for $\beta = \gamma = 0.0006$ it is $t_{osc} \approx 1200$, for $\beta = \gamma = 0.001$ it is $t_{osc} \approx 730$, and for $\beta = \gamma = 0.0015$ it is $t_{osc} \approx 580$. These data agree very well with the occurrence of the first minimum of $\sigma$ in dependence on $\tau$ as presented in Fig.~\ref{fig8}. Accordingly, we also conclude that the information transmission delay induced transitions to spatiotemporal synchronization of neuronal activity are due to the locking between $\tau$ and the predominant oscillation period of individual neurons on the scale-free network.

\section{Summary and discussion}

In sum, we have studied front propagation and synchronization transitions on scale-free neuronal networks in dependence on the information transmission delay, coupling strength and the average degree. We found that an appropriately adjusted delay length can significantly enhance synchronization of excitatory fronts in an intermittent fashion in dependence on $\tau$. The intermittent outlay emerges due to the locking between the delay and the inherent oscillation frequency of individual neurons on the scale-free network. Thus, approximately at every multiple of the inherent oscillation period of each neuron, the information transmission delay between coupled neurons results in supremely ordered and synchronized fronts of excitations. The widths of these dynamical regimes in dependence on $\tau$ can be broadened, and the synchronization further improved, if the coupling strength or the average degree of the network is enlarged. In addition, we have examined the robustness of these findings to different levels of additive noise, as well as to different system sizes and scaling exponents characterizing the scale-free topology, finding that all the conclusions prevail. We have shown that fine-tuned information transmission delays can effectively supplement some recently identified mechanisms for the enhancement of synchronization \cite{mh1, mh2}, as well as weak signals in general \cite{mh3}, on scale-free networks. These conclusions seem to be supported by real biological data stating that conduction velocities along axons connecting neurons vary from 20 to 60 m/s \cite{TS1}. Real-life transmission delays are within the range of milliseconds, suggesting that substantially lower or higher values may be preclusive for optimal functioning of neuronal tissue. In future studies, it would be interesting to examine the impact of synaptic noise and different conductance states \cite{oz1} on synchronization transitions in delayed complex networks, as well as to pinpoint the precise role of different aspects of structure and functioning of active neuronal networks \cite{bj1}. We hope that our present study will be a useful source of information when striving towards these goals.

\begin{acknowledgments}
This work was supported by the National Science Foundation of China (Funds Nos. 10702023 and 10832006) and China's Post-Doctoral Science Foundation (Funds Nos. 200801020 and 20070410022). Matja{\v z} Perc additionally acknowledges support from the Slovenian Research Agency (Grant Z1-2032-2547).
\end{acknowledgments}

\end{document}